\documentclass[12pt]{IEEEtran}
\usepackage[pdftex]{graphicx}
\usepackage[outdir=./]{epstopdf}
\usepackage{subfigure}

\usepackage{threeparttable}
\usepackage{tabularx}
\newcolumntype{Y}{>{\centering\arraybackslash}X}
\usepackage{booktabs}
\usepackage{multirow} 
\usepackage[export]{adjustbox}

\expandafter\let\csname equation*\endcsname\relax
\expandafter\let\csname endequation*\endcsname\relax
\usepackage{amsmath}

\usepackage[acronym]{glossaries}
 \usepackage{array,multirow,graphicx}
\newacronym{PSG}{PSG}{polysomnography}
\newacronym{HRV}{HRV}{heart rate variability}
\newacronym{SD}{SD}{standard deviation}
\newacronym{ANS}{ANS}{autonomic nervous system}
\newacronym{REM}{REM}{rapid eye movement}
\newacronym{AASM}{AASM}{American Association of Sleep Medicine}
\newacronym{RnK}{R\&K}{Rechtschaffen and Kales}
\newacronym{PSQI}{PSQI}{Pittsburgh Sleep Quality Index}
\newacronym{PPG}{PPG}{photoplethysmography}
\newacronym{EEG}{EEG}{electroencephalography}
\newacronym{EOG}{EOG}{electrooculography}
\newacronym{EMG}{EMG}{electromyography}
\newacronym{LSTM}{LSTM}{long- and short-term memory}
\newacronym{ECG}{ECG}{electrocardiography}
\newacronym{ICD-10}{ICD-10}{International Classification of Diseases, tenth edition}
\newacronym{IBI}{IBI}{inter-beat interval}
\newacronym{VLF}{VLF}{very low frequency}
\newacronym{LF}{LF}{low frequency}
\newacronym{HF}{HF}{high frequency}
\newacronym{DFA}{DFA}{detrended fluctuation analysis}
\newacronym{CRF}{CRF}{conditional random fields}

\usepackage{todonotes}

\begin{document}

\title{LSTM knowledge transfer for HRV-based sleep staging}

\author{
M Radha$^{124}$, P Fonseca$^{124}$, A Moreau$^{3}$, M Ross$^{3}$, A Cerny$^{3}$, P Anderer$^{3}$, R M Aarts$^{24}$
\thanks{$^1$ Joint first authorship}
\thanks{$^2$ Personal Health, Philips Research, Royal Philips}
\thanks{$^3$ Sleep and Respiratory Care, Royal Philips}
\thanks{$^4$ Electrical Engineering, Eindhoven University of Technology}
\thanks{Correspondence: mustafa.radha@philips.com}}
\maketitle

\begin{abstract}
Automated sleep stage classification using \gls{HRV} is an active field of research. In this work  limitations of the current state-of-the-art are addressed through the use of deep learning techniques and their efficacy is demonstrated. First, a temporal model is proposed for the inference of sleep stages from \gls{ECG} using a deep \gls{LSTM} classifier and it is shown that this model outperforms previous approaches which were often limited to non-temporal or Markovian classifiers on a comprehensive benchmark data set (292 participants, 541214 samples) comprising a wide range of ages and pathological profiles, achieving a Cohen's $\kappa$ of $0.61\pm0.16$ and accuracy of $76.30\pm10.17$ annotated according to the \gls{RnK} annotation standard.\\ 
Subsequently, it is demonstrated how knowledge learned on this large benchmark data set can be re-used through transfer learning for the classification of \gls{PPG} data. This is done using a smaller data set (60 participants, 91479 samples)  that is annotated with the more recent \gls{AASM} annotation standard, achieving a Cohen's $\kappa$ of $0.63\pm0.13$ and accuracy of $74.65\pm8.63$ for wrist-mounted \gls{PPG}-based sleep stage classification, higher than any previously reported performance using this sensor modality. This demonstrates the feasibility of knowledge transfer in sleep staging to adapt models for new sensor modalities as well as different annotation strategies.

\end{abstract}


%
%
%
%
%

\section{Introduction}
Sleep is a reversible state of disconnection from the external environment characterized by reduced vigilance and quiescence. It plays an essential role in the diurnal regulation of mind and body in mammals, and is hypothesized to have a wide array of functions ranging from digestion to memory consolidation. The objective measurement of sleep in adult humans involves sleep staging: the process  of segmenting a sleep period into \textit{epochs}, typically 30 seconds long, and assigning a sleep stage to each epoch. The \gls{AASM} \cite{berry2012aasm} distinguishes five sleep stages: \gls{REM} sleep, three levels of non-REM sleep (NREM1, NREM2, NREM3) and wake (W). Sleep staging is usually performed by visually scoring the electric activity in the brain, eye movement and chin muscles, measured respectively with \gls{EEG}, \gls{EOG} and \gls{EMG}. Together with sensors measuring cardiac and respiratory activity, this setup is collectively referred to as \gls{PSG}. \\

Although it remains the gold-standard for clinical assessment of sleep and diagnosis of sleep disorders, \gls{PSG} is practically limited to one or two measuring nights, and cannot be effectively performed at home for a prolonged period of time. With the increasing availability of inexpensive and robust physiological sensors, cardiac and respiration-based sleep stage classification have been actively researched over the last few years, as they might provide a highly unobtrusive and cost-efficient surrogate of \gls{PSG} sleep scoring. \\

Cardiac sleep stage classification in particular makes uses of the well-known relation between sleep stages and \gls{ANS} activity, usually measured with different \gls{HRV} characteristics, or features. The inference of sleep stages is done by training machine learning algorithms which translate \gls{HRV} features to sleep stages. Usually, a set of  physiological features make up the feature space $\mathcal{X}$, with a marginal probability distribution $P(X)$. Together they form the domain $\mathcal{D} = \{\mathcal{X}, P(X)\}$ of the sleep staging problem. The sleep stage label space $\mathcal{Y}$ then, in the simplified case of four-class sleep staging, comprises  the labels $W, N1/2, N3, R \in \mathcal{Y}$ (corresponding to Wake, combined N1 and N2, N3 and REM sleep) and the  conditional distribution $P(Y|X)$. Finally, the classification task is a combination of the label space distribution and the domain: $\mathcal{T} = \{\mathcal{Y}, P(Y|X)\}$. Models are trained with training sample pairs $x \in X$ and $y \in Y$ for each 30 second segment of sleep (i.e. epoch).\\ 

\subsection{Cardiac features}
\label{sec:intro:hrv}
Although \gls{HRV} features can be derived from \gls{ECG}, this is arguably not the most convenient sensor for prolonged monitoring at home. Sensors such as wrist-worn reflective \gls{PPG} are more adequate for the purpose of long-term sleep monitoring, since they are more comfortable, and can be easily set up by the individuals under investigation. However, the use of these sensors outside the consumer domain remains limited and very few data sets exist with which sleep staging classifiers can be adequately trained. In particular for models with many free parameters the size of the training set is likely to play a crucial role. While a model can be trained on \gls{HRV} features derived from readily-available \gls{ECG}, it is likely to perform sub-optimally for \gls{PPG}-derived \gls{HRV} for three reasons: coverage is typically lower; beat localisation methods are slightly less accurate than their \gls{ECG} counterparts; and the time delay between the heart contraction (\gls{ECG} R-peak) and the arrival of the pulse at the wrist \gls{PPG}, known as the pulse transit time, is not constant: it is continuously modulated through properties of the arterial vessels such as blood pressure and vasoconstriction \cite{radha2017arterial}, which change throughout the night \cite{radha2018_bp_dipping}.\\

This motivates the need to develop techniques that on the one hand can make use of larger data sets with \gls{ECG}, while on the other can adapt to \gls{PPG}. This is not only true for \gls{PPG}, there are many other unobtrusive sensors for which such a methodology would yield better algorithms, such as ballistocardiography, galvanic skin response or stationary camera-based \gls{PPG}. \\

\subsection{Sleep stage annotation}
\label{sec:intro:annotation}
Visual sleep stage scoring with \gls{PSG} is usually performed with a fixed set of rules and guidelines. The current standard is maintained by the \gls{AASM} \cite{berry2012aasm}. However, before 2007 the \gls{RnK} \cite{Rechtschaffen1968} guidelines were the most commonly used, since its publication in 1968. The main difference is that the older standard distinguishes between 4 classes of non-REM sleep, namely S1, S2, S3 and S4 (where S3 and S4 have been merged by the \gls{AASM} into a single stage, NREM3). Next to that, and over the last few years, additional refinements and rules were introduced in the \gls{AASM} manual to improve the inter-rater agreement between expert annotators. These small changes have been shown to lead to different results. It was found that there was a structural increase in NREM1 and NREM3, while NREM2 prevalence decreased. For REM, there seemed to be an age-dependent difference in annotation, with less REM being scored in younger people \cite{moser2009sleep}. Although the change in scoring guidelines aimed to improve the overall scoring quality, it created a disparity between the annotations in older and newer data sets. When re-scoring of large, older data sets is impractical or prohibitively expensive, they might be of limited use. This motivates the need for machine learning methods that can exploit the large similarity between \gls{RnK} and \gls{AASM} guidelines, while adapting sufficiently to the current standard and future changes. 

\subsection{Machine learning algorithms}
\label{sec:intro:ml}
The most popular algorithms in sleep stage research, which we have also used in the past \cite{Radha2014,fonseca2015sleep}, are feed-forward models in which $P(Y_t | X_t)$ is estimated, where $Y_t$ and $X_t$ are respectively the label and features at time step $t$. Given that sleep architecture has common temporal patterns throughout the night, this approach may not achieve optimal performance as it does not exploit the dependency between time steps. To this end, approaches such as Elman recurrent networks \cite{hsu2013automatic}, hidden Markov models \cite{mendez2010sleep} and \gls{CRF} \cite{fonseca2017cardiorespiratory} were proposed, in which the conditional distribution is made dependent on the previous time point: $P(Y|X_t, X_{t-1})$. These models strongly outperform earlier models, demonstrating the value of modeling temporal dependence in sleep stage classification \cite{fonseca2018comparison}. Given this improvement in performance, these approaches motivate the investigation of better temporal models that can take into account more than only the last time step, especially since it was shown in earlier work \cite{garcia2016probabilistic} that transition probabilities of sleep stages in a Markov model are different for each sleep cycle, i.e. that $P(Y_t|Y_{t-1})$  changes throughout the night, violating the Markov assumption for the linear chains traditionally used, and emphasizing the need for a non-Markovian temporal model.\\

\subsection{Objectives}
In this work a new methodology for automated cardiac sleep staging will be presented that can leverage the long-term contextual information in sleep, building upon recent advancements in \gls{LSTM} algorithms and deep neural networks. Next to that, a transfer learning methodology will be evaluated for the adaptation from \gls{ECG} to \gls{PPG} features as well as from \gls{RnK} to \gls{AASM} annotations as a means to learn domain- and target- invariant knowledge from older, larger \gls{ECG} data sets and re-use the knowledge to train an algorithm using \gls{PPG} and the \gls{AASM} standard using smaller, newer data sets.

\section{Materials and Methods}
\label{sec:materials_methods}
\subsection{Materials}
\label{sec:materials}
\subsubsection{Siesta data set}
The first data set used in this study was collected as part of the EU SIESTA project \cite{klosh2001siesta} in the period from 1997 to 2000 in seven European countries. The study was approved by the local ethical committee of each research group and all participants signed informed consent. Participants had no history of alcohol or drug use or worked shifts. The data set includes 195 healthy participants and 97 patients with a sleep or sleep-disturbing disorder (26 patients with insomnia, 51 with sleep apnea, 5 with periodic limb movement disorder and 15 with Parkinson's disease) \cite{klosh2001siesta}.

Each participant was monitored for a total of 15 days and at day 7 and 8 participants were invited to sleep in the sleep laboratory to collect overnight \gls{PSG}. Each recording was scored by two trained somnologists from different sleep centers according to the \gls{RnK} guidelines \cite{Rechtschaffen1968}, and revised by a third expert who took the final decision in case of disagreement. More details regarding participants and study design were described in \cite{klosh2001siesta}. Table~\ref{table:demographics} indicates the participant demographics, and corresponding sleep statistics.\\

\subsubsection{Eindhoven data set}
The second data set was collected in 2014 and 2015 in Eindhoven, the Netherlands, approved by the Internal Committee of Biomedical Experiments of Philips Research and conducted in accordance with the Declaration of Helsinki. All participants gave informed consent before participation.  It includes 101 recordings of 60 healthy participants with no primary history of neurological, cardiovascular, psychiatric, pulmonary, endocrinological, or sleep disorders. In addition, none of the participants were using sleep, antidepressant or cardiovascular medication, recreational drugs or excessive amounts of alcohol. 
Each of the participants underwent one or two nights of \gls{PSG} measurements in a hotel, where in addition to the standard montage recommended by \gls{AASM} for offline sleep scoring \cite{berry2012aasm}, a CE-marked logging device containing a \gls{PPG} and a tri-axial accelerometer sensor (Royal Philips, Amsterdam, the Netherlands) was used. The logging device was mounted on the non-dominant wrist of the participant, with the sensor facing the skin on the dorsal side of the hand, above the ulnar styloid process. The \gls{PSG} data was annotated by a trained sleep technician according to the \gls{AASM} rules of sleep scoring \cite{berry2012aasm}. More details regarding study design were described in \cite{fonseca2017validation}. Table~\ref{table:demographics} indicates the participant demographics, and sleep statistics for this data set.\\

\begin{table*}[!t]
	\centering
	\footnotesize
	\begin{threeparttable}
		\caption{\label{table:demographics}Demographics and sleep statistics of participants in the two data sets used in the study. Sleep statistics are computed based on the sleep stage annotation of the data set.}
		\begin{tabular*}{ 0.9 \textwidth}{l @{\extracolsep{\fill}} cccc}
				\toprule
				& \multicolumn{2}{c}{Siesta} & \multicolumn{2}{c}{Eindhoven}  \\
				\cmidrule{2-3} \cmidrule{4-5} 
				Parameter       & Mean (SD) & Range                                  & Mean (SD) & Range \\
				\midrule
				N               & \multicolumn{2}{c}{292 participants, 584 recordings} 	 & \multicolumn{2}{c}{60 participants, 101 recordings}  \\
				\multirow{2}{*}{Sex} 
				                & \multicolumn{2}{c}{126 female participants (43.2\%)}   & \multicolumn{2}{c}{26 female participants (43.3\%)}  \\	
		                        & \multicolumn{2}{c}{252 female recordings (43.2\%)}  & \multicolumn{2}{c}{48 female recordings (47.5\%)} \\	
				Age (year)	    &  $ 51.5 (17.3) $ &  $ 20.0-95.0 $                  &  $ 51.1 (7.9) $ &  $ 41.0-66.0 $  \\ 
				BMI (kg/m$^2$)  &  $ 25.6 (4.5)  $ &  $ 16.5-43.3 $                  &  $ 25.6 (3.9) $ &  $ 17.5-36.2 $  \\ 
				TIB (hour)	    &  $ 8.0 (0.5)   $ &  $ 5.8-9.6   $                  &  $ 7.9 (0.7)  $ &  $ 6.4-10.3  $  \\ 
				SE (\%)		    &  $ 80.8 (12.8) $ &  $ 14.6-99.1 $                  &  $ 85.0 (9.8) $ &  $ 36.0-96.6 $  \\ 
				N1 (\%)		    &  $ 13.1 (8.4)  $ &  $ 2.4-77.1  $                  &  $ 10.7 (5.0) $ &  $ 3.0-30.6  $  \\ 
				N2 (\%)		    &  $ 53.8 (8.8)  $ &  $ 13.6-78.8 $                  &  $ 41.7 (8.7) $ &  $ 22.2-66.6 $  \\ 
				N3 (\%)		    &  $ 13.8 (8.4)  $ &  $ 0.0-44.5  $                  &  $ 26.2 (8.7) $ &  $ 10.3-47.3 $  \\ 
				REM (\%)	    &  $ 18.2 (5.9)  $ &  $ 0.0-34.8  $                  &  $ 21.4 (5.9) $ &  $ 9.2-38.2  $  \\ 
				\bottomrule				
			\end{tabular*}
		\begin{tablenotes}[para,flushleft]
			\footnotesize
			N1, N2, N3, and REM percentages were calculated over the total sleep time for each recording.\\
			BMI: body mass index, TIB: time in bed, SE: sleep efficiency.
		\end{tablenotes}
	\end{threeparttable}
\end{table*}

\subsection{Feature extraction}
This study used a set of 135 \gls{HRV} features extracted from \glspl{IBI} computed from either \gls{ECG} or \gls{PPG}. All features were described in earlier work, \cite{Fonseca2016,fonseca2018comparison} and are only summarized here. 
To extract features from \gls{ECG}, the signal was first high-pass filtered to remove baseline wander using a Kaiser window of 1.016 s, a cut-off frequency of 0.8 Hz and a side-lobe attenuation of 30 dB. QRS complexes were detected with a Hamilton-Tompkins QRS detector \cite{hamilton2002open,hamilton1986quantitative} and further localized with a post-processing algorithm \cite{Fonseca2014}. \Glspl{IBI} were then computed from the time difference between pairs of consecutive R-R peaks.
To extract features from \gls{PPG}, individual heart beats were first extracted from the pulsatile component of the PPG using the same algorithm as used by Papini et al. \cite{Papini2017}. \Glspl{IBI} were then computed from the time difference between pairs of consecutive pulses.

From the resulting \gls{IBI} time series, features describing \gls{HRV} characteristics were computed using windows of 4.5 min centered on each 30 second epoch of the recording. The time domain features included the mean absolute distance, mean, median, 5th, 10th, 25th, 75th, 90th and 95th percentiles, percentage of successive \glspl{IBI} longer than 50 ms, range, root mean square of successive \gls{IBI} differences, standard deviation, and standard deviation of successive \gls{IBI} differences, all calculated on the original and on the detrended \cite{Redmond2006} \gls{IBI} and instant heart rate time series \cite{Redmond2006,Fonseca2016}. The frequency-domain \gls{HRV} features included the normalized PSD of a 4 Hz-interpolated \gls{IBI} series in the \gls{VLF}, \gls{LF} and \gls{HF}, the LF-HF ratio, and the peak frequency and power in the \gls{HF} band, using fixed \cite{Busek2005} and adapted boundaries \cite{Long2014}, and the maximum phase and module in the HF pole \cite{mendez2010sleep}. The \gls{HRV} feature set further included the mean, median, minimum and maximum posterior probabilities of arousals given sequences of five consecutive beats on each 30 second epoch \cite{basner2007ecg}. \Gls{DFA} was applied on the \gls{IBI} time series to calculate the scaling exponent for faster, slower, and all time scales, \cite{Kantelhardt2001,penzel2003comparison} and the scaling exponent was also computed using Peng's method \cite{Peng1994}. In addition, windowed \cite{Adnane2012} and progressive \cite{Staudacher2005} \gls{DFA} were computed. The multiscale sample entropy of \gls{IBI} time series was calculated at length 1 and 2, and scales 1–10 over windows of 510 s \cite{Costa2002} in addition to the sample entropy of symbolic binary changes in consecutive \glspl{IBI} \cite{Cysarz2000}.\\

Furthermore, Teager energy was used to characterize transition points and local maxima in \gls{IBI} series \cite{Kvedalen2003}, including the mean energy, percentage of transition points and maxima, mean and standard deviation of intervals between transition points and maxima, mean and standard deviation of the amplitude of normalized \glspl{IBI} at transition points and maxima, all calculated based on the \gls{IBI} time series, and on the first intrinsic mode function after empirical mode decomposition \cite{huang1998empirical}.\\

To express the interaction between cardiac and respiratory autonomic activity, a cardiac-to-respiratory phase synchronization ratio was computed in terms of percentage of beats synchronized with a 6:2, 7:2, 8:2, 9:2 and the dominant ratio, short- and long-term coordination in terms of presence and duration of synchronized heart beats \cite{cysarz2004quantitative,Bettermann2000} and Higuchi's fractal dimension measure of phase coordination \cite{Higuchi1988}. Finally, visibility graphs were used to model cardiorespiratory interaction in the \gls{IBI} series, and used to calculate the assortativity mixing coefficient, the mean and standard deviation of the clustering coefficients and degrees, slope of the power-law fit to the degree distribution and percentage of nodes with a small and with a high degree, all computed based on the visibility and difference visibility graphs.

\subsection{Long- and short term memory network}
In this work bi-directional multi-level \gls{LSTM} networks \cite{hochreiter1997long} are proposed as solution to the temporal processing limitations of models commonly used for this task, as described in Section \ref{sec:intro:ml}. \glspl{LSTM} consist of memory cells that can store long-term information from time series and  generate an output based on the current time step input, their last output  (short-term recurrence) and the internal cell state  (long-term recurrence). The cell state is controlled through gating mechanisms. A detailed description and equations of \gls{LSTM} cells are given in the original paper \cite{hochreiter1997long}. Stacking multiple layers of \gls{LSTM} cells allows for the memorization of deeper temporal structures in the data. By having two \gls{LSTM} stacks in parallel, one applied in the forward and another in the backward direction, it is possible to take into account both past and future data to classify each single time step \cite{graves2005bidirectional}. The proposed architecture is a neural network consisting of three distinct blocks: 


\paragraph{Domain layer} consists of perceptrons that make 32 linear combinations of features. The function of this layer is to pre-weight, pre-select and combine the input into a more compact representation of the domain $\mathcal{D}' = \{\mathcal{X}',P(X')\}$
\paragraph{Temporal layer} consists of the \gls{LSTM} stacks. These \glspl{LSTM} take $\mathcal{D}'$ over all time steps and generate 128 new features at each time step, where temporal information has been taken into account through the short- and long-term recurrence properties both from the past and future. This results in yet another translation of the domain $\mathcal{D}" = \{\mathcal{X}",P(X")\}$.
\paragraph{Decision layer} consists of two levels of perceptrons, of which the final level generates four sigmoid outputs representing the class probabilities $P(Y| X")$, given $\mathcal{D}"$ at that time step. The outputs sum up to one for each classification through softmax normalization.

\begin{figure}
\centering
\includegraphics[scale=0.33]{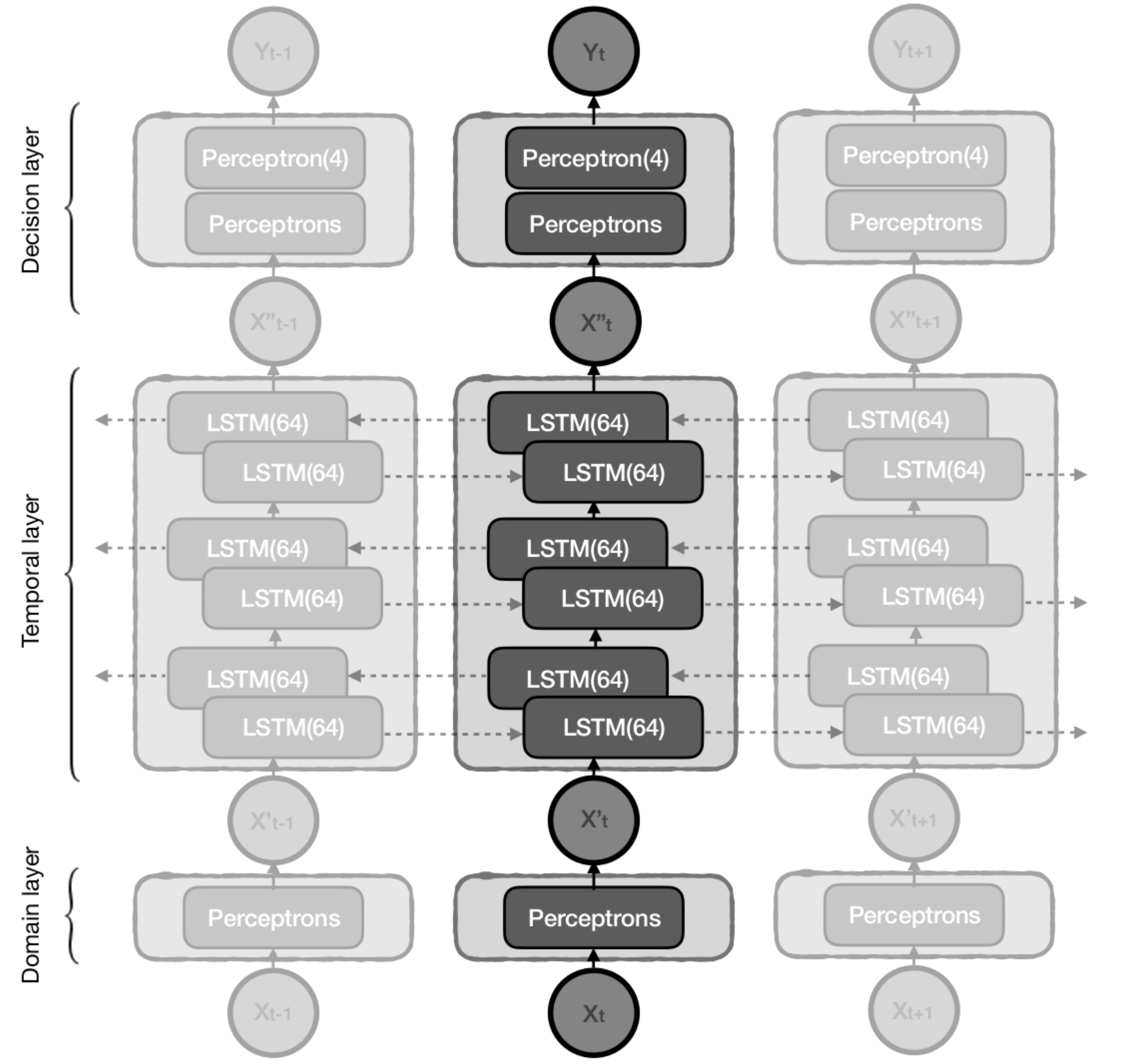}	
\caption{Overview of the neural network architecture.}
\label{fig:lstm}
\end{figure}

The complete model is illustrated in Figure \ref{fig:lstm}. The $2.6 \cdot 10^5$ free parameters of the model were trained simultaneously with the RMSprop optimizer,a variant of AdaDelta \cite{zeiler2012adadelta} introduced in a lecture series \cite{tieleman2012lecture}. Dropout \cite{srivastava2014dropout} on the input (20\%), on \gls{LSTM} outputs (50\%) and recurrent connections (50\%) was applied during training to reduce overfitting. Categorical cross-entropy was used as the loss function during model fitting.

\subsection{Model training}
\subsubsection{Siesta model}
The model will be initially trained and evaluated on the Siesta database using \gls{HRV} derived from \gls{ECG} ($X_{S,ecg}$, where $S$ denotes Siesta), which will be named the \textit{Original} model. This will be done using a 75-25\% training-validation split of the complete Siesta data set which will allow an assessment of the overall performance of the \gls{LSTM} architecture on a larger set of \gls{ECG}-based \gls{HRV} data.\\ 

While the model can be trained using $X_{S,ecg}$, it is likely to not be optimal for \gls{PPG}-derived \gls{HRV} ($X_{ppg}$) data due to the differences between \gls{PPG} and \gls{ECG} (as explained in Section \ref{sec:intro:hrv}). These differences are apparent in the Eindhoven data: the correlation between $X_{E,ecg}$ and $X_{E,ppg}$ (where $E$ denotes Eindhoven) is $0.77\pm0.15$. In addition, Eindhoven data is annotated with the new \gls{AASM} standard which might be a further source of performance degradation of the Original model (trained on \gls{RnK} annotation), as described in Section \ref{sec:intro:annotation}. The Original model will be evaluated on the Eindhoven data: using $X_{E,ecg}$ to show the performance difference due to annotation strategy, and $X_{E,ppg}$ to show the further difference due to the different sensor modality.\\

\subsubsection{Eindhoven model}
An alternative approach is to directly train the model on $X_{E,ppg}$ and $X_{E,ecg}$. However the model might reach a sub-optimal performance given the large amount of free parameters and the limited size of the Eindhoven data. This will be shown through 4-fold cross-validation on the Eindhoven data for both \gls{ECG} and \gls{PPG}. 

\subsubsection{Transfer learning}
To overcome the above issues, transfer learning is proposed, which involves transferring model knowledge to solve a new but related problem where less data samples are available. This technique has been proven effective in the context of deep neural networks, where knowledge is represented in a  modular structure of layers and has been applied successfully in both computer vision \cite{Sun2016} and natural language processing \cite{huang2013cross} networks. Transfer learning is practically achieved by retraining a small portion of a large model that had been previously trained on a larger data set.\\

In this case, the hypothesis is that the \gls{LSTM} layer contains generalizable knowledge with respect to domain and target. This layer also contains 96.6\% of the model's weights, making it a good candidate for transfer. Thus the candidate layers for retraining are the domain and decision layers. Retraining the domain layer would imply adapting the domain layer to make $X'$ more comparable to what the Siesta model expects (see Figure \ref{fig:lstm}). However, this operation alone may not be sufficient, as fixing subsequent layers will leave little room for optimization.  An alternative is to retrain only the decision layer. In this case, the Siesta model will be used to extract $X"$ from Eindhoven data, but the final aggregation step will cope with changes due to sensor modality and annotation strategy. The third option is to retrain both domain and decision layers simultaneously, leaving it up to the optimizer to change weights as needed.\\ 

The models created with these three transfer learning strategies will be referred to as the \textit{Domain}, \textit{Decision} and \textit{Combined} models (referring to which layers are retrained), and their performance on the Eindhoven data will be statistically compared against (1) the original trained model on the Siesta data set, which will be referred to as \textit{Siesta model}, and (2) the model trained directly on Eindhoven data without prior training, which will be called the \textit{Eindhoven model}.\\ 

These approaches will be evaluated using $X_{E,ecg}$ to show how much the transfer can adapt to the new \gls{AASM} annotation strategy; and $X_{E,ppg}$ to show how much the model can adapt to both annotation strategy and the \gls{PPG} modality. All transfer learning experiments will be performed in 4-fold cross-validation on the Eindhoven data set. \\ 

\subsubsection{Evaluation methodology}
All training-validation and fold splits are done such that all recordings of a given participant are part of the training or the validation groups, but not both, or of a single fold in the case of cross-validation.  The same fold assignments are used for all cross-validation experiments in the Eindhoven data set, to allow for a paired comparison of the validation performance. Performance is evaluated using metrics of accuracy and Cohen's $\kappa$ coefficient of agreement on an epoch-per-epoch basis in comparison with ground-truth. Differences in performance were compared on a per recording basis using a paired two-tailed Wilcoxon signed-rank test.

\section{Results}
The results of training the Siesta model on the Siesta data set using \gls{ECG} and \gls{RnK} annotations are presented in Table~\ref{table:training_performance_siesta}, where the average and the standard deviation of the performance for both the training and validation performance are shown. The evaluation of the model using the Eindhoven data set as hold-out set (without transfer learning) is shown in Table~\ref{table:training_performance_siesta} both for the \gls{ECG}-derived as well as for the \gls{PPG}-derived \gls{HRV} features. A significant difference ($p < .001$) was found between the training and the validation performance on Siesta. The Non-transfer strategy resulted in a $\kappa$ of $0.55\pm0.14$ and accuracy of $69.82\%\pm10.23\%$ for \gls{PPG} while for \gls{ECG} a $\kappa$ of $0.60\pm0.12$ and accuracy of $72.97\%\pm9.08\%$ was found. These results were not significantly different from the performance of the Siesta model on the same data.\\

\begin{table*}[!t]
	\centering
	\footnotesize
	\begin{threeparttable}
		\caption{\label{table:training_performance_siesta}Performance of the Siesta model on the training and validation parts of Siesta as well as on the Eindhoven data set using ECG and PPG.}
		\begin{tabularx}{\textwidth}{l YYY}
			\toprule
			
				 Data set                & Recs. (subjs.) & $\kappa$ (-)    & Accuracy (\%)     \\
				 \midrule
				 Siesta Training split   & 440 (220)      & $0.69 \pm 0.11$ & $82.03 \pm 5.89$  \\
				 Siesta Validation split & 148 (74)       & $0.61 \pm 0.16^1$ & $76.30 \pm 10.17^1$ \\
				 Eindhoven ECG           & 101 (60)       & $0.62 \pm 0.11$ & $74.60 \pm 7.66$  \\
				 Eindhoven PPG           & 101 (60)       & $0.57 \pm 0.12$ & $71.88 \pm 8.34$  \\
				
			\bottomrule
		\end{tabularx}
			\begin{tablenotes}[para,flushleft]
				\footnotesize
				$^1$Significant difference ($p < .001$) found between Training and Validation performance, after a Wilcoxon rank-sum test.\\
			\end{tablenotes}
	\end{threeparttable}
\end{table*}

\begin{figure}
    \centering
    \begin{subfigure}
        \centering
        \includegraphics[width=7cm,valign=t]{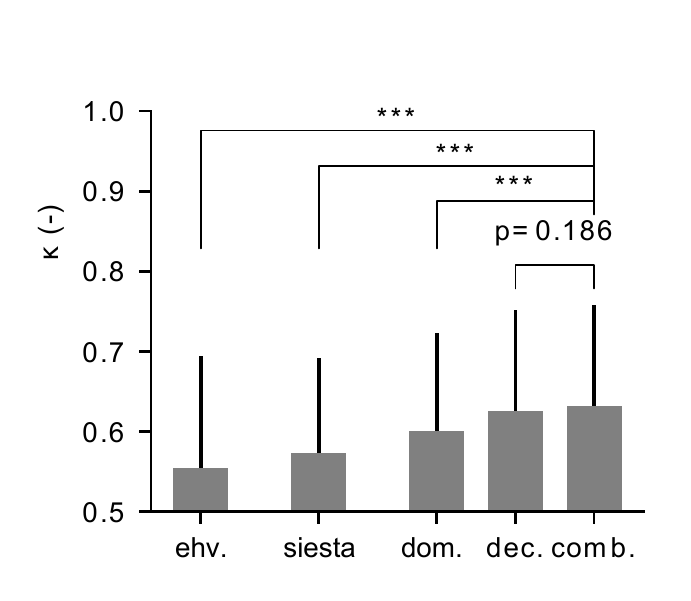}
    \end{subfigure}\\
    \begin{subfigure}
        \centering
        \includegraphics[width=7cm,valign=t]{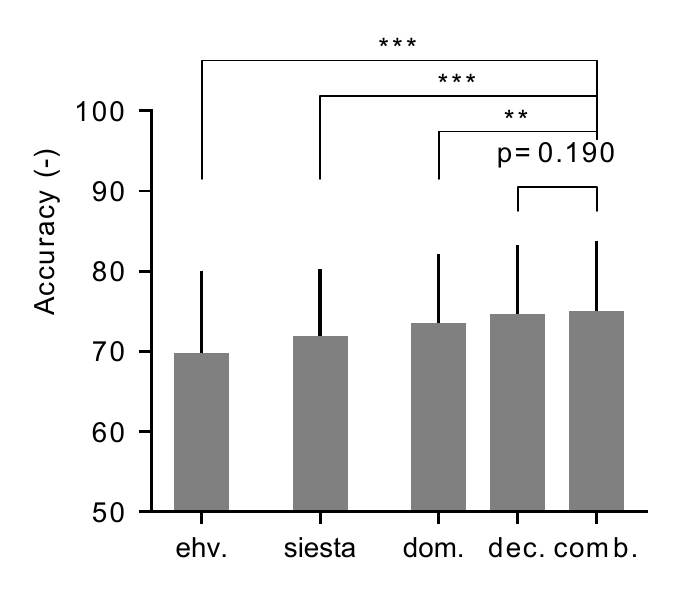}
    \end{subfigure}
    \caption{\label{fig:statistical_comparisons_ppg}Performance comparisons between the validation results obtained with Combined (`comb.'), Domain (`dom.'), Decision (`dec.'), Siesta ('siesta') and Eindhoven (`ehv.') training strategies on the Eindhoven data set using PPG. $^{\ast\ast\ast}$ and $^{\ast\ast}$ indicate significant differences (at $p < .001$ and $p<.01$ respectively) after a Wilcoxon signed rank test.}
\end{figure}

\begin{figure}[t]
    \centering
    \begin{subfigure}
        \centering
        \includegraphics[width=7cm,valign=t]{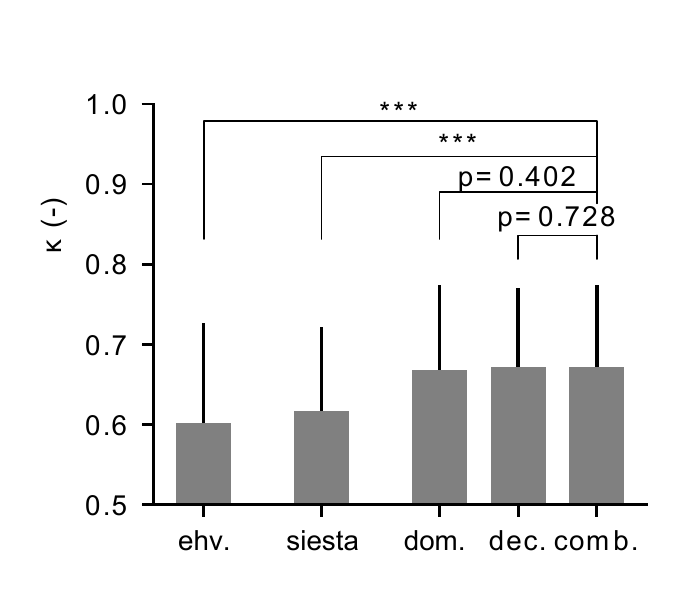}
    \end{subfigure}\\
    \begin{subfigure}
        \centering
        \includegraphics[width=7cm,valign=t]{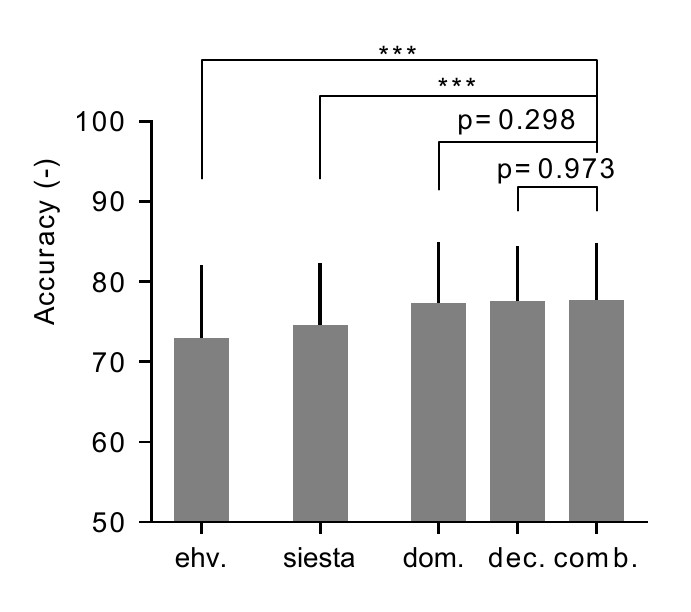}
    \end{subfigure}
    \caption{\label{fig:statistical_comparisons_ecg}Performance comparisons between training strategies on the Eindhoven data set using ECG. See caption to Figure~\ref{fig:statistical_comparisons_ppg}.}
\end{figure}

The performance after adapting the Siesta model using the three transfer experiments on the Eindhoven data set was derived for both \gls{ECG} and \gls{PPG} (separately) and shown in Table \ref{tab:transfer_performance}. These metrics were compared with the performance of both Siesta and Eindhoven model performances. Results of these comparisons are illustrated in Figure~\ref{fig:statistical_comparisons_ppg} for PPG and Figure~\ref{fig:statistical_comparisons_ecg} for ECG. For PPG, the performance obtained with Combined is different from all strategies except Decision for both $\kappa$ and accuracy (statistically significant). For ECG, the performance obtained with Combined is only significantly different from cross-validation and Siesta model performance. 

\begin{table*}
	\centering
	\footnotesize
	\begin{threeparttable}
		\caption{Cross-validation results on Eindhoven data set with transfer learning. Domain, Decision and Combined refer respectively to the models where the domain, decision and both of them simultaneously are adapted.}
		\label{tab:transfer_performance}
		\setlength\tabcolsep{1.5pt} 
		\begin{tabularx}{\textwidth}{l YY c YY}
			\toprule
								&  \multicolumn{2}{c}{PPG}           & &  \multicolumn{2}{c}{ECG}          \\
			\cmidrule{2-3} \cmidrule{5-6} 
					      & $\kappa$        & Accuracy (\%)    & & $\kappa$        & Accuracy (\%)    \\
			 \midrule
			 Domain   & $0.60 \pm 0.12$ & $73.53 \pm 8.60$ & & $0.67 \pm 0.11$ & $77.34 \pm 7.62$ \\
			 Decision & $0.63 \pm 0.13$ & $74.65 \pm 8.63$ & & $0.67 \pm 0.10$ & $77.51 \pm 6.98$ \\
			 Combined & $0.63 \pm 0.13$ & $75.04 \pm 8.72$ & & $0.67 \pm 0.10$ & $77.64 \pm 7.23$ \\
			\bottomrule		
		\end{tabularx}
	\end{threeparttable}
\end{table*}

\section{Discussion}
\subsection{\gls{LSTM} as a sleep stage classifier}
\label{disc:lstm}
The level of agreement between an \gls{HRV}-based sleep staging algorithm and \gls{PSG} for the \gls{LSTM} model (Table \ref{table:training_performance_siesta}) is beyond anything reported before in a data set containing a wide age range (20-95 years old) and a number of sleep disorders, including sleep apnea, insomnia, Parkinson's disease and a few participants with periodic limb movement disorders. As reported in our earlier work on \gls{HRV}-based sleep stage classifiers \cite{fonseca2018comparison}, the best performance achieved in a comparable part of the Siesta data  using a non-temporal model (Bayesian Linear Discriminant) yielded $\kappa=0.40$ and accuracy of 59.98\% for 4-class sleep stage classification. When temporal models such as \gls{CRF} \cite{fonseca2017cardiorespiratory} were used, this performance rose to a $\kappa$ of 0.47 and accuracy of 68.77\%, still well below the performance of the \gls{LSTM} model presented in this study.\\ 

There are a few studies that have achieved a performance closer (but not equal) to the current study, though those were limited to a smaller set comprising only healthy ($\kappa$ of $0.56$ \cite{fonseca2015sleep}) and/or young participants ($\kappa=0.56$ \cite{willemen2014evaluation} and $0.60$ \cite{zhao2017learning}, which are known to be easier to classify with \gls{HRV} features \cite{fonseca2018comparison}. A closer analysis of the results obtained for the validation split of Siesta (Table~\ref{table:training_performance_siesta}) revealed that for individuals below 45 years old the average $\kappa$ was 0.67, higher than for middle-aged individuals (between 45 and 65, with a $\kappa$ of 0.64) and much higher than for elderly (above 65, with a $\kappa$ of 0.45), with a significant ($p<0.001$) correlation between age and $\kappa$ of $-0.55$. In addition, all studies discussed in this paragraph used, besides \gls{HRV}, also respiration as an input modality, which was not used in the current study.\\
 
Interestingly, two recent studies on the use of neural networks for \gls{EEG}-based sleep stage classification \cite{Stephansen2017,Anderer2018} also achieved superior performance using \gls{LSTM} in their classification models. Given the state-of-the-art, it could be argued that \gls{LSTM} models and other similar non-linear temporal classifiers adequately represent temporal relations during sleep and could be a key ingredient in future sleep staging algorithms.

\subsection{Transfer learning}
The average performance of the Siesta-trained model, when used with $X_{E,ppg}$, dropped from a $\kappa$ of 0.61 to 0.57. However, the performance using $X_{E,ecg}$ remained approximately the same with a $\kappa$ of 0.62.  While it is clear that the model degrades when used with \gls{PPG}, the expected degradation on the \gls{ECG} due to a change in scoring standard might have been offset by the fact that the Eindhoven data set only contains healthy participants which are, in general, slightly easier to classify using \gls{HRV} than sleep-disordered patients \cite{fonseca2018comparison}).

\subsubsection{Adaptation to \gls{PPG}}
Both  the Combined and Decision adaptation methods improved $\kappa$ to an average of 0.63 for \gls{PPG} as compared to 0.57 before transfer, while accuracy improved  from 72\% to 75\%. The Domain approach resulted in a model that adapted halfway: $\kappa$ increased to 0.60 and accuracy to 74\%. Since there was no significant difference in performance for the Decision and Combined training strategies, we speculate that adapting the final Decision layer only would be sufficient, with the Combined approach not adding any extra value, while the Domain strategy proving to be insufficient for this task.\\ 

The performance obtained for \gls{PPG} is considerably higher than that reported in state-of-the-art. Only two previous validation studies for \gls{PPG}-based sleep staging are known to the authors, and which reported a $\kappa$ of 0.42 \cite{fonseca2017validation} and 0.52 \cite{Beattie2017}. Both studies used only healthy participants and in addition made use of an accelerometer to improve the detection of the Wake class. With a $\kappa$ of 0.55, training \gls{LSTM} directly on the Eindhoven \gls{PPG} data (without transfer) already achieved a higher performance than reported in these studies. After transfer learning the performance sets a considerably higher standard for \gls{PPG}-based sleep stage classification, bringing this convenient sensing modality a step closer to the gold-standard \gls{PSG}. \\

\subsubsection{Adaptation to \gls{AASM}-based scoring}
A similar performance improvement was gained when doing transfer learning to $X_{E,ecg}$ as to $X_{E,ppg}$. This could imply that part of the performance improvement obtained with transfer for both \gls{ECG} and \gls{PPG} is due to adaptation to \gls{AASM} standard rather than to the sensor modality; this is compatible with the fact that the \gls{AASM} scoring rules are more consistently reproducible, resulting in a higher inter-rater agreement between annotators \cite{danker2009interrater}, even though the Siesta data set was annotated by consensus according to \gls{RnK} rules rather than a single annotator using \gls{AASM} rules. The relatively small amount of adaptation to \gls{PPG} could be attributed to the fact that \gls{PPG} is a lower quality sensor of \gls{HRV}: lower coverage means that model can only learn to deal with missing data, but not infer knowledge that it could not have learned from \gls{ECG}. However it should be noted that for the \gls{PPG} adaptation, it was not sufficient to only transfer the domain layer (see Figure 
\ref{fig:statistical_comparisons_ppg}) while for the 
\gls{ECG} transfer this approach did not differ from the other strategies (see Figure 
\ref{fig:statistical_comparisons_ecg}). This would suggest that the adaptation to \gls{PPG} requires a larger optimisation space due to the change in modality. This could be related to differences between \gls{PPG} and \gls{ECG} that are not related to coverage, such as the variable time delays due to pulse transit time (see Section \ref{sec:intro:hrv}).\\

A side-effect that was not hypothesised, is that the model might have also learned \gls{HRV} characteristics specific of the Eindhoven data set, which comprises exclusively middle-aged healthy participants, as opposed to the Siesta data set which comprises, besides healthy participants, participants with sleep disorders as well as a wider range of ages. However, the Eindhoven data set does not include participants younger than 40 years old, which as was explained in Section \ref{disc:lstm} are easier to classify than middle-aged and older individuals. Thus, it is unlikely that the population characteristics are the only driver of the performance gain after transfer, though untangling and quantifying the effects of these factors should be a topic of future work.\\

\section{Conclusion}
In conclusion, a new method has been proposed for automated \gls{HRV}-based sleep staging consisting of an \gls{LSTM} neural network. The network has a higher capacity to learn and exploit temporal characteristics of sleep data than previous non-temporal or Markovian models and resulted in a increased performance in the Siesta data set when compared with previous work, showing its adequacy for participants of all ages and different pathological conditions.  The second objective was to investigate transfer learning methodologies to adapt this model to a new problem and target domain, namely for the classification of \gls{HRV} data from \gls{PPG} instead of \gls{ECG}, as well as to adapt to \gls{AASM} guidelines for sleep staging instead of the older \gls{RnK} guidelines. It was shown that the use of transfer learning resulted in a better model for this problem in comparison with a complete training of the model from scratch on this (smaller) data set. The performance achieved for \gls{PPG} in this data set is considerably higher than reported in earlier work on \gls{PPG}-based sleep scoring. The examined transfer strategies in most cases did not differ from each other, however for the transfer to \gls{PPG} the decision layer was required to be part of the transfer for the best results. \\

It remains to be conclusively elucidated which factors contribute most to the performance improvements, but evidence suggests that the transfer to the modern \gls{AASM} guidelines, which have been shown to yield higher inter-rater agreement, might be an important cause. Future work should focus on understanding whether similar improvements can be achieved with \gls{PPG} sleep data of disordered participants, as well as replicating the transfer experiment to other unobtrusive modalities such as in-bed ballistocardiographic sensors.

\bibliographystyle{IEEEtran}
\bibliography{library}

\end{document}